# 基于实时路况地图的养护作业开始时间优化

许哲谱，杨群

（同济大学道路与交通工程教育部重点实验室，上海市 201804）

**摘要**：对养护作业开始时间进行优化可以极大地减少养护作业活动给出行者造成的延误。本研究首先提出了一种基于实时路况地图的实时路况数据采集方法，然后提出可以将实时路况转换成实时交通量的方法，基于此实时交通量数据和经典的基于排队论的延误计算参数法，可以计算不同养护作业开始时间下的延误，进而得到最优养护作业开始时间。通过实际案例验证了本文提出的实时路况到实时交通量的转换方法的可靠性以及基于实时路况地图数据对养护作业开始时间进行优化的可行性。

**关键词**：养护作业; 时间优化; 实时路况地图; 数据采集; 排队论

**中图分类号**：U418.2 **文献标志码**：A

## Road Maintenance Operation Start Time Optimization Based on Real-time Traffic Map Data

*Zhepu Xu, Qun Yang*

(The Key Laboratory of Road and Traffic Engineering, Ministry of Education, Tongji University, Shanghai, 201804)

**Abstract:** Optimizing the maintenance operation start time can greatly reduce the delays caused by the maintenance operations. A real-time traffic status data acquisition method based on real-time traffic map was first proposed, and then a method that can convert real-time traffic status into real-time traffic volume was put forward. Based on this real-time traffic volume data and the classic delay calculation method based on queuing theory, the delays caused by maintenance operations at different start time can be calculated and compared, and therefore the optimal maintenance operation start time can be obtained. The feasibility of the real-time traffic status data to real-time traffic volume data conversion method and the feasibility of optimizing the maintenance operation start time based on real-time traffic map data are verified by actual cases.

随着汽车保有量的增加、道路上实际运行的车辆数量越接近道路通行能力，道路就越容易形成拥堵。道路上的拥堵大致可以分为常发性拥堵和非常发性拥堵，而人们最无法忍受的是那种无法预知的非常发性拥堵，其给出行者带来无法预期的延误。养护作业正是导致这种无法预期延误的重要因素之一[1]。

为了尽可能减少养护作业活动对大众出行的影响，人们做了很多工作。其中养护作业开始时间优化非常重要：在最佳的时间开始作业，可以尽可能减少延误。养护作业开始时间优化的关键问题是延误的计算，延误计算方法大致可以分为三类：参数法、非参数法和仿真方法[2,3]。

基于排队论的延误计算方法是最常用的参数法[4,5]。该方法的关键参数主要有交通量、养护作业区通行能力和养护持续时间等。由于这些关键参数比较容易获得，人们对该方法研究较多，这个方法也已被广泛应用到实践中。Abraham 和

Wang[6]于 1981 年、Dudek 和 Richards[7]于 1982 年、Morales[8]于 1986 年、Schonfeld 和 Chien[9-11]于 1999 年、2001 年和 2002 年分别用确定性排队模型对作业区内的车流排队延误进行了研究。此后，又不断有研究人员对该模型进行完善，Chen 和 Schonfeld 于 2006 年[12]，Tand 和 Chien 于 2008[13]，Meng 和 Weng 于 2013 年[14]分别针对现有模型的不足之处作出了改进。以排队论为代表的的参数法的优点是原理简单，操作方便，在数据量不充分的情况下也能保证一定的准确性，不足之处是对现实因素的考虑较简单，相比于其他方法精度不是特别高。

非参数法的思想是基于大量的数据，建立养护作业因素、道路因素等与延误之间的关系模型。相比于参数法有明确的表达式，非参数法的模型往往是复杂而隐藏的，比如神经网络模型等。Du 和 Chien 于 2016 年采用多层前馈人工神经网络模型，建立了养护作业区基础数据、探测车数据与延误之间的关系[2]。2017 年，Du 和 Chien 又提出了一个支持向量机+ANN 的复合机器学习模型对 2016 年模型进行改进，进一步提高延误计算精度[15]。非参数法的优点是能够获取更高的精度，缺点是对数据要求高，需要获取到大量且完整的历史数据。

仿真方法就是利用宏观、中观或者微观交通仿真工具，仿真养护作业的存在所带来的延误。比如，2009 年 Hsin-Yun Lee 运用 VISSIM 交通仿真软件对高速公路施工作业区施工安排和施工速度控制进行仿真，研究成果可以降低 11%交通延误[16]。2010 年朱永光利用 VISSIM 微观交通软件分析了公路养护施工区的通行能力的影响因素、作业段长度与道路使用者延误的关系等[17]。仿真方法的优点是仿真模型一旦标定好后可以用于研究各种养护条件对通行延误的影响，不足之处是要建立一个好的仿真模型需要有完善的历史交通数据，仿真模型的标定过程非常耗时，仿真对计算资源占用很大，仿真时间也比较长。

可以看到，在养护作业延误分析方面有一定的研究基础，现有的三种方法，即参数法、非参数法和仿真方法都有各自的优点和不足。其中参数法凭借着其简单易操作，对数据要求相对较低的特点有很大的运用场景，特别是在养护作业数据量不足的情况下可以为养护决策和规划提供一定的参考。该方法的关键参数之一是交通路况数据。常用的路况数据采集方法有[18]：（1）人工采集法，即派人在道路现场进行采集；（2）基于固定设施的自动采集法，即通过在道路上铺设线圈、安装摄像头、测速雷达等固定设施，实现路况的自动采集；（3）浮动车法，在出租或公交上安装 GPS 获取车辆的位置与速度信息，并及时将数据提交给智能交通系统。

路况数据采集方法的不断丰富，给我们提供了更多的便利。然而养护事件可能发生在道路的任意位置，也就是说养护决策最理想的状态是掌握道路任意位置的交通状况。这一点采用人工采集和固定式交通路况检测方法不可行，而采用基于浮动车法所采集的数据又都集中在少数商业公司，养护部门很难获取。因此考虑其他间接方法获取道路任意位置的实时路况数据很有现实意义。在这一方面，国内外研究有过一些尝试：P.

Pokorny 利用谷歌地图开发接口，编程实现了一个实时路况获取系统，用于评价城市实时交通水平[19]。刘瑶杰基于百度地图，获取了北京市 5 环以内 7 天的实时路况数据，并基于这些数据挖掘城市交通拥堵的时空分布特征，探索交通拥堵产生和发展的地理规律[20]。王芹则基于高德地图，也采用类似的手段获取实时交通大数据，研究城市交通拥堵热点判别方法[21]。

随着数据获取技术的进步，基于实时路况地图获取实时交通量数据可以为参数法准备更充分的数据基础，能够更好的发挥参数法的优势。本文基于商业地图的实时路况数据，采用基于排队论的参数法构建养护作业开始时间优化模型，弥补现有方法的不足，以期减少养护所造成的非常发性拥堵对出行者的影响。

接下来介绍本文剩余内容的安排：首先介绍本文提出的基于实时路况地图的养护作业开始时间优化方法，包括实时路况数据的获取、实时路况数据到实时交通量的转换、基于排队论的延误计算以及开始作业时间优化方法等；然后用实际案例验证本文方法的有效性并演示本文方法在实践中的应用；最后对本文做一个总结和展望。

# 1 方法

养护作业开始时间优化的关键是延误的计算，本文所采用的延误计算方法是基于排队论的参数法。而延误计算的关键是道路交通量的获取和养护作业区通行能力的估计，这也是本方法的主要内容。如图 1 所示，方法由四个步骤组成：（1）基于实时路况地图获取交通量，（2）养护作业区通行能力估计，（3）基于排队论计算延误，（4）养护作业开始时间优化。其中第一步基于实时路况地图获取交通量又可以划分为三个小步骤：采集与积累实时路况数据，将实时路况数据定量化，然后将实时路况转换成交通量。由于基于排队论的延误分析方法已经是比较成熟的方法，而关于延误分析所需要的通行能力估计也已经有大量研究，本研究的难点就在于获取实时路况数据，并得到实时交通量数据，因此这部分重点介绍步骤一的内容。

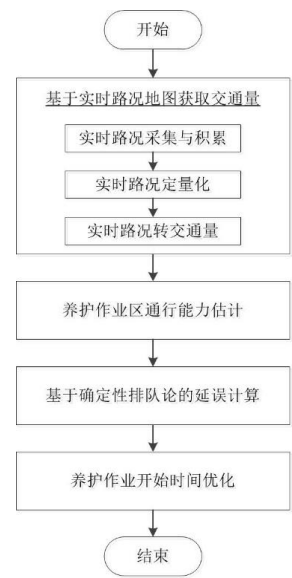

图 1 本文方法主要流程

Fig.1 Main method of this paper

## 1.1 实时路况采集与积累

实时路况是电子地图（如百度地图）常见的功能，用不同的颜色将道路的实时拥堵情况反映出来，方便用户出行参考。根据实时路况数据粒度的粗细程度，实时路况地图通常采用 3-5 种颜色不等。典型的，用绿色表示畅通、橙色表示缓行、红色表示拥堵、紫色表示严重拥堵（彩色实时路况地图样例见附件 A 图 A-1）。

实时路况采集与积累要实现的效果是，输入

某地理位置坐标，可以监测该位置的实时路况，从而达到采集和积累实时路况数据的目的。根据各商业地图提供的调用接口，实现上述目的大概可以分为构建参数、获取地图切片和路况状态识别三个步骤：

（1）构建参数，即构建调用接口需要的参数。其中，关键的是坐标之间的转换。整个过程会涉及到五种坐标的转换：地理坐标系统、平面坐标系统、像素坐标系统、图块坐标系统和可视区域坐标系统，即将所要监测位置的地理坐标转换成可视区域坐标，关于坐标转换的详细介绍可以参考[22]。

（2）获取地图切片，即调用商业地图开放的接口，获取第一步中传入参数所对应的切片。以百度地图为例，获取切片的地址为http://its.map.baidu.com:8002/traffic/TrafficTileService?time=1527043432323&label=web2D&v=016&level=19&x=105113&y=27854，其中 x 和 y 就是步骤（1）中构建的坐标参数，彩色实时路况地图切片样例见附件 A 图 A-2。

（3）路况状态识别，即识别切片上可视区域坐标位置的交通状态，可以通过像素的 RGB 进行识别，比如，对于百度地图的状态，可以采用表 1 的值进行识别。

表 1 交通状态与 RGB 范围对照表
Table 1 Traffic status and corresponding RGB range

| 实时交通状态 | 颜色 | RGB 阈值 |
|---|---|---|
| 顺畅 | 绿色 | $G \neq B\ AND\ R \leq 240$ |
| 缓行 | 橙色 | $G \neq B\ AND\ R > 240$ |
| 拥堵 | 红色 | $G = B\ AND\ R \geq 200$ |
| 严重拥堵 | 紫色 | $G = B\ AND\ R < 200$ |

有了上述基础，可以实现输入一个指定的坐标，就获得该位置实时交通状态的效果。通过设置定时器，可进一步实现按照一定频率监测指定位置的交通状况。

由于养护事件可能会发生在道路的任意位置，因此必须实现能够自动采集道路任意位置的交通状况。考虑到道路发生交通拥堵时都是以段落的形式呈现的，即一个指定点的交通状态可以从它周边的点的交通状态估计，因此可以先将道路网离散成点网，然后运用上述基于位置的交通状态采集手段监测所有的点，达到监测道路网的效果。具体步骤如下：

第一步，采用线转点工具将道路网离散成点网。常用的 GIS 软件都有线转点工具，需要注意的是，为了兼顾后续数据监测的效率和养护实际需求，同一路段上点之间的距离不应太大也不宜太小。经过实践以及参考现有研究[20,21]，可以取 50m 作为离散粒度。

第二步，获取所有道路点的坐标，这里需要注意不同坐标系的转换。

第三步，将基于位置的交通状态自动采集方法运用于上述各个点，实现道路点网的日常交通状态数据自动采集。

## 1.2 实时路况定量化

通过 1.1 实时路况数据采集系统采集到的是交通的状态量（比如顺畅、缓行、拥堵和严重拥堵）。实时路况地图是地图公司融合浮动车数据、道路固定监测装置采集的数据、人工采集数据等得到的。研究表明，实时路况数据与车辆平均速度有很强的相关性[23,24]，因此可以通过标定来建立路况与车辆平均速度的对应关系，从而对实时

路况进行定量化。主要思想是用 1.1 实时路况采集系统采集某位置处的实时路况数据，同时在现场采集车辆速度数据，通过分析两套数据的统计特征得到实时路况数据和速度的对应关系，之后只要用路况采集系统采集到了交通状态，就能估计出当时的平均速度。为了保证得到完整的实时交通状态与速度的对应关系，在数据采集位置选择时应该选择容易发生拥堵的位置，以保证采集到各个状态的数据。具体的操作将会在第 2 部分以实例的形式加以演示。

### 1.3 实时路况转交通量

通过 1.2 的定量化过程，将交通状态量转化成了速度，而延误计算需要的是交通量，因此需要将速度转换成交通量，这很容易让人想起经典的速度-流量模型，即在连续交通流中，速度 U、流量 V 和密度 K 之间存在着如下关系[25]：

$$V = UK.$$

通常假设密度 K 与流量 V 呈线性关系，便可推导得速度 U 与流量 V 的二次抛物线关系模型，如图 2 及式(1)所示。

$$V = K_j(U - \frac{U^2}{U_0}) \tag{1}$$

式中：$K_j$ 是阻塞密度；$U_0$ 是零流量时的平均车速。

然而，该模型有一个很大的局限性是不能预测交通量大于通行能力时（V/C>1）的车辆行驶速度。理论上说，当交通量大于通行能力时，路段交通阻塞，此时，即使到达车辆数增加，能通过的交通流量仍只能是通行能力，即路段流量不能大于通行能力，剩余车辆数(到达车辆数 V-通行能力 C）会排队等候。但养护作业延误分析仍需要预测当路段上的交通需求量(即车辆到达数)超过通行能力时，这些车辆的平均行驶速度，但经典模型却无能为力了[26]。

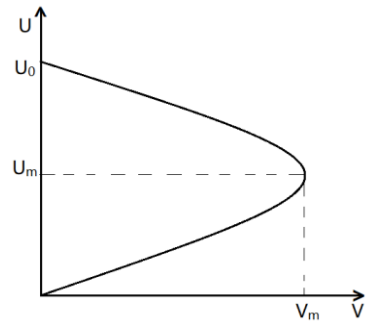

图 2 U-V 典型图式

Fig.2 Classic relationship between U and V

早在 2003 年，王炜就提出了一个实用车速-流量关系模型[26]，可以弥补经典模型的不足，下面简单介绍该实用模型。

#### 1.3.1 道路交通流车速-流量实用关系模型

实用关系模型认为车速-流量模型应该是 S 型曲线，其数学表达式为：

$$\left. \begin{array}{l} U = \dfrac{\alpha_1 U_s}{1+(\dfrac{V}{C})^\beta} \\ \beta = \alpha_2 + \alpha_3(\dfrac{V}{C})^3 \end{array} \right\} \tag{2}$$

式中，$\alpha_1$，$\alpha_2$，$\alpha_3$ 为回归参数；$U_s$ 是设计车速（km·h$^{-1}$)，在有实测自由流速度时，宜采用自由流速度。

表 2 列出了高速公路车速-流量实用模型参数，其他等级道路参数参考原论文[26]。

由式(2)确定的高速公路不同设计车速下的车速-流量曲线如图 3 所示。

**表 2 高速公路车速-流量通用模型参数表**

Table 2 Parameters of the practical model for expressways

| 设计车速$U_s$ (km·h$^{-1}$) | 单车道通行能力 C(pcu·h$^{-1}$) | $\alpha_1$ | $\alpha_2$ | $\alpha_3$ |
|---|---|---|---|---|
| 120 | 2200 | 0.93 | 1.88 | 4.85 |
| 100 | 2200 | 0.95 | 1.88 | 4.86 |
| 80 | 2000 | 1.00 | 1.88 | 4.90 |
| 60 | 1800 | 1.20 | 1.88 | 4.88 |

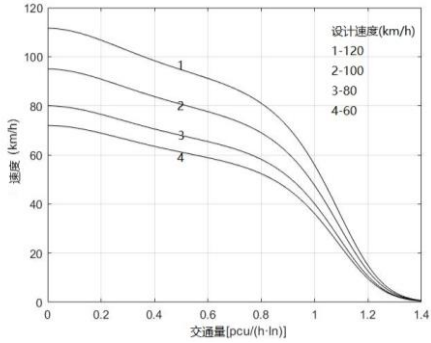

图 3 高速公路实用车速-流量曲线

Fig 3. Practical U-V model for expressways

### 1.3.2 正常道路通行能力估计

实用模型的一个重要参数是正常道路通行能力，这已经有成熟的方法和文献可供参考。以高速公路基本路段通行能力为例，参考《公路通行能力手册》[27]，可通过如下步骤确定其通行能力。

首先确定高速公路基本路段的基准通行能力，手册已经给出，比如对于基准自由流速度为80km·h$^{-1}$时，基准通行能力值为1800pcu·(h·ln)$^{-1}$。

实际道路条件往往达不到基准条件，需要对自由流速度进行修正，按式(3)进行：

$$v_{FF} = v_{BFF} + \Delta v_w + \Delta v_N \quad (3)$$

式中：$v_{FF}$是高速公路基本路段实际自由流速度(km·h$^{-1}$)；$v_{BFF}$是基本路段的基准自由流速度(km·h$^{-1}$)；$\Delta v_w$是车道宽度和路侧宽度对基准自由流速度的修正值(km·h$^{-1}$)；$\Delta v_N$是车道数对基准自由流速度的修正值(km·h$^{-1}$)。参数的具体取值参考手册。

那么，实际道路交通条件下的通行能力可以按式(4)计算：

$$C_r = C_b \times f_{HV} \times f_p \times N \quad (4)$$

式中：$C_r$是实际道路、交通条件下单车道的通行能力(pcu·h$^{-1}$)；$C_b$是基准条件下五级服务水平（V/C=1）对应的单车道基准通行能力(pcu·h$^{-1}$)；$f_{HV}$是交通组成修正系数；$f_p$是驾驶人总体特征修正系数；$N$是单向车道数。参数具体取值参考手册。

### 1.4 养护作业区通行能力估计

排队论计算延误的另一个重要参数是养护作业区的通行能力，这也可以参考现有的道路通行能力手册进行估计。实际通行能力可以按照以下公式（5）计算[27]。

$$C_{rs} = C_{bs} \times f_n \times f_{lw} \times f_{lc} \times f_{HV} \\ \times f_{se} \times f_{wi} \times f_{ls} \quad (5)$$

式中：$C_{rs}$是施工区实际通行能力[pcu·(h·ln)$^{-1}$]；$C_{bs}$是施工区基准通行能力[pcu·(h·ln)$^{-1}$]；$f_n$是车道数修正系数；$f_{lw}$是车道宽度修正系数；$f_{lc}$是侧向净空修正系数；$f_{HV}$是交通组成修正系数；$f_{se}$是限制速度修正系数；$f_{wi}$是作业强度修正系数；$f_{ls}$是光照条件修正系数。各参数的具体取值参考手册。

### 1.5 养护作业区延误计算

养护作业区的交通延误可以分为四个部分

[28,29]：（1）进入作业区范围后，车辆由正常速度降低到作业区要求的速度所造成的的减速延误；（2）车辆以比较低的速度通过作业区造成的运行延误；（3）车辆通过作业区后恢复到正常速度时的加速延误；（4）作业区范围内的排队延误。

这四部分交通延误中，前三项延误可以比较车流通过作业区的速度和正常路段时的速度得到。第四项排队延误，对于拥挤状态有此项延误很容易理解，因为拥挤状态下车流会在作业区内排队；但是非拥挤状态也可能会形成排队，这是由于车流到达的随机性，在道路流量没超过作业区通行能力的部分时段内，也会出现排队现象。排队延误可以用排队论计算，但是非拥挤状态和拥挤状态的排队延误计算方法有一定的差异。

现假定车流在正常路段上的行驶速度为$V_1$，在作业区段的行驶速度为$V_2$，在驶入作业区时的减速度和加速度分别为$a_1$和$a_2$，作业区长度为$L$，作业总时间为$T$，作业区的通行能力为$C$，假定每个数据采集小间隔$t_d$内达到作业区的交通流是均匀的，为$Q_i$。根据$t_d$时间段内车辆累积情况以及车辆到达交通量和通行能力的关系划分成两级判断条件，那么每个$t_d$内单个车辆的减速延误$d_1$，运行延误$d_2$，加速延误$d_3$，排队延误$d_4$，所有车辆的排队延误$D_4$，以及车辆总延误$DL_i$计算公式如表3所示，具体的推导过程参考文献[29]。

那么，养护作业导致的总延误就是所有$DL_i$的总和。值得注意的是，日常情况下道路也可能会发生拥堵产生延误，本文所说的延误特指由养护作业造成的额外延误。

### 1.6 养护作业开始时间优化

通过前面部分的介绍，可以计算以给定时刻作为养护作业开始时间的延误，那么很容易想到，可以遍历一天中的任意时刻作为养护开始时间，分别计算延误，将使得延误最小的开始时间作为最优开始时间。

表3 作业区每个$t_d$的总延误$DL_i$计算公式汇总

Table 3 Calculation formulas for delay in each period

| 一级判断 | 二级判断 | 当前时段延误计算公式 | 备注 |
|---|---|---|---|
| 当前无累积排队车辆 | $Q_i \leq C$ | $d_1 = (V_1 - V_2)^2 / (2 \cdot a_1 \cdot V_1)$<br>$d_2 = (1/V_2 - 1/V_1) \cdot L$<br>$d_3 = (V_1 - V_2)^2 / (2 \cdot a_2 \cdot V_1)$<br>$d_4 = Q_i / [C \cdot (C - Q_i)]$<br>$DL_i = (d_1 + d_2 + d_3 + d_4) \cdot Q_i \cdot t$ | 此时路况畅通 |
| | $Q_i > C$ | $d_1 = (V_1 - V_2)^2 / (2 \cdot a_1 \cdot V_1)$<br>$d_2 = (1/V_2 - 1/V_1) \cdot L$<br>$d_3 = (V_1 - V_2)^2 / (2 \cdot a_2 \cdot V_1)$<br>$D_4 = Q_{i-1} t_d + \frac{1}{2}(Q_i - C) t_d^2$<br>$DL_i = (d_1 + d_2 + d_3) \cdot Q_i \cdot t + D_4$ | 排队开始形成 |
| 当前有累积排队车辆 | $Q_i < C$ | $d_1 = (V_1 - V_2)^2 / (2 \cdot a_1 \cdot V_1)$<br>$d_2 = (1/V_2 - 1/V_1) \cdot L$<br>$d_3 = (V_1 - V_2)^2 / (2 \cdot a_2 \cdot V_1)$<br>$D_4 = \frac{Q_{i-1}^2}{2(C - Q_i)}$<br>$DL_i = (d_1 + d_2 + d_3) \cdot Q_i \cdot t + D_4$ | 排队开始消散 |
| | $Q_i \geq C$ | $d_1 = (V_1 - V_2)^2 / (2 \cdot a_1 \cdot V_1)$<br>$d_2 = (1/V_2 - 1/V_1) \cdot L$<br>$d_3 = (V_1 - V_2)^2 / (2 \cdot a_2 \cdot V_1)$<br>$D_4 = Q_{i-1} t_d + \frac{1}{2}(Q_i - C) t_d^2$<br>$DL_i = (d_1 + d_2 + d_3) \cdot Q_i \cdot t + D_4$ | 排队继续增加 |

## 2 验证与应用

这一部分将以实例的方式对前述方法进行验证以及应用，主要分为两部分内容：（1）实时路

况定量化和速度-流量实用模型的验证；（2）养护作业区延误计算以及开始时间优化。

## 2.1 实时路况定量化以及实用模型的验证

2.1.1 数据采集

数据采集主要分为三个部分，即实时路况采集、交通量采集和速度采集。实时路况利用 1.1 节所述的采集系统进行自动采集，确定采集位置后系统按照一分钟一次的频率采集所在位置的交通状态。本实例中，实时路况地图采用谷歌地图。交通量和速度数据采用现场架设摄像机录制交通流视频，后期通过视频分析的方式获取。

为了采集到畅通、缓行、拥堵和严重拥堵四个状态下的交通数据，本次数据采集选择了苏黎世 E60 高速某双向 4 车道局部路段，该路段会发生常发性拥堵。

如图 4 所示，采用两台摄像机分别在 A 处和 B 处录制交通视频，其中 A 处是易发生拥堵的地方，B 处在 A 的上游 2 公里处，B 处基本上处于畅通状态。录制视频的同时，利用实时路况采集系统采集该路段的实时路况数据。

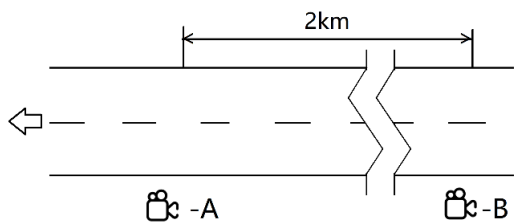

图 4 数据采集地点示意图

Fig.4 Illustration of the data collection

2.1.2 实时路况定量化

通过交通流录像分析，得到车辆的速度，并结合实时路况数据，对不同路况的车辆速度进行分析，拟合成正态分布，可以得到图 5 所示速度分布图。取各状态曲线的交点作为各状态的分界点，并取各拟合正态曲线的均值作为该状态的定量指标，如表 4 所示。其中，速度范围是地图公司状态划分标准在实际道路上的表现情况[24]。

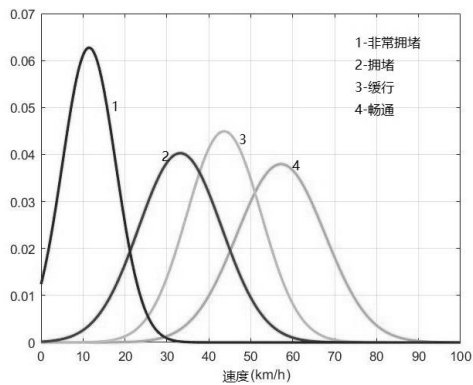

图 5 E60 速度分布图

Fig 5. Speed distribution of E60

表 4 E60 实时路况状态定量表

Table 4 Quantification values for the real-time traffic status of E60

| 路况 | 畅通 | 缓行 | 拥堵 | 严重拥堵 |
|---|---|---|---|---|
| 速度范围($km·h^{-1}$) | 51-90 | 38-51 | 21-38 | 0-21 |
| 定量化值（$km·h^{-1}$） | 57 | 44 | 33 | 12 |

2.1.3 实用模型验证

由于这一部分验证案例中采集到了实际的速度和交通量数据，因此可以根据实际数据得到真实的通行能力和自由流速度，否则在没有条件的情况下就只能采用手册推荐的值。

通过视频数据提取，获取不同时间段，不同路况状态下的交通量和对应的速度，并按照经典的速度-流量抛物线模型（式 1）进行拟合，可以得到如图 6 所示的数据分布图，$R^2$=0.96。可以把最大流率 1577$pcu·(h·ln)^{-1}$ 作为该路段的通行能力，自由流速度为 91$km·h^{-1}$。

根据 1.3.1 所述的模型，以及上述实测数据，对于本路段，取 $α_1$=1，$α_2$=1.88，$α_3$=4.90，$U_s$=91

可以得到如图 7 所示的实用速度-流量曲线。

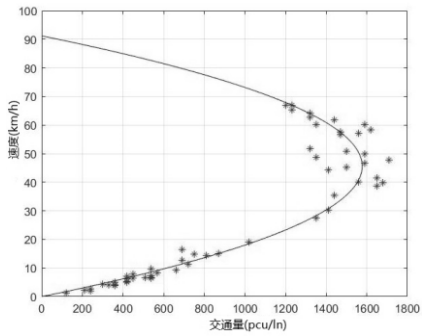

图 6 速度-流量拟合

Fig 6. Classic U-V model of E60

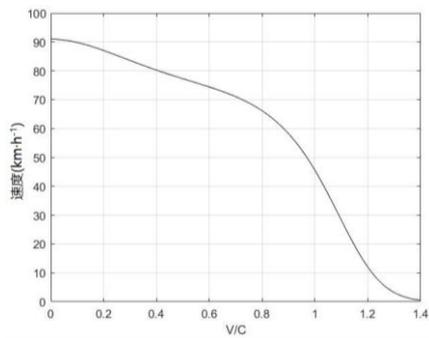

图 7 实例的实用速度-流量曲线图

Fig 7. Practical U-V model of E60

由图 7 可见，按照该模型的估计，当 v/c 为 0.88 即交通量为 1388pcu·(h·ln)⁻¹ 时，速度是 60km·h⁻¹ 左右。通过视频处理，找到交通量为该值且交通状态属于畅通部分的速度（表 5）。可见，本速度-流量模型在 v/c<1 的情况下对速度和流量的估计是比较准确的。从图 7 也可以看到，当交通量达到路段通行能力时(即 v/c=1)，路段车辆速度的平均值在 45km·h⁻¹ 左右，对应着实时交通路况的缓行状态。也就是说路段的通行能力可以用路段交通量从不拥堵到拥堵变化的交通量表示，这个观点正是文献[30]所提出的路段通行能力界定方法。最后，验证该模型在 v/c>1 情况下的适用性。显然，该情况一定出现在实时路况为拥堵或者严重拥堵状态的时间段。由表 4 可知，拥堵状态下，车辆的平均速度在 33 km·h⁻¹ 左右，由图 7 可得对应的 v/c 为 1.07，即交通量为 1688pcu·(h·ln)⁻¹。根据经典的速度-流量曲线，当道路拥堵时,其交通量甚至小于路段通行能力，实际数据采集时也发现确实存在这个现象，所以通过测量拥堵路段交通量的方法并不能得到交通需求量（即车辆到达数），这必须通过计算路段上游不拥堵区段车辆到达数来实现。选取上游处于畅通状态而下游处于拥堵路段的时间段统计车辆达到数，得到表 6 所示的数据。可见在道路拥堵情况下，交通流量确实达到了 1688pcu·(h·ln)⁻¹，从而验证了实用模型的适用性。

表 5 v/c=0.88 时的速度和流量数据（部分）

Table 5 Speed and volume data when v/c=0.88(part)

| 序号 | 交通量 [pcu·(h·ln)⁻¹] | 速度均值 (km·h⁻¹) | 序号 | 交通量 [pcu·(h·ln)⁻¹] | 速度均值 (km·h⁻¹) |
| --- | --- | --- | --- | --- | --- |
| 1 | 1500 | 64.4 | 5 | 1440 | 61.8 |
| 2 | 1320 | 64.0 | 6 | 1350 | 60.3 |
| 3 | 1560 | 63.3 | 7 | 1470 | 57.6 |
| 4 | 1320 | 62.7 | 8 | 1470 | 56.6 |

表 6 上游交通量数据（部分）

Table 6 Upstream volume data(part)

| 序号 | 交通量 [pcu·(h·ln)⁻¹] | 序号 | 交通量 [pcu·(h·ln)⁻¹] | 序号 | 交通量 [pcu·(h·ln)⁻¹] |
| --- | --- | --- | --- | --- | --- |
| 1 | 1800 | 7 | 1650 | 13 | 1590 |
| 2 | 1770 | 8 | 1650 | 14 | 1590 |
| 3 | 1710 | 9 | 1620 | 15 | 1590 |
| 4 | 1710 | 10 | 1620 | 16 | 1590 |
| 5 | 1680 | 11 | 1620 | 17 | 1590 |
| 6 | 1680 | 12 | 1590 | | |

最后，可以对照一下实时路况状态和实际采集的路况状态。根据定量化表（表 4）和上述实用模型（图 7）可以得到，当交通状态为畅通即速度为 57km·h⁻¹ 时，交通量为 1453 pcu·(h·ln)⁻¹；当交通状态为缓行即速度为 44km·h⁻¹ 时，交通量

为 1593pcu·(h·ln)⁻¹；当交通状态为拥堵即速度为 33km·h⁻¹ 时，交通量为 1687pcu·(h·ln)⁻¹；当交通状态为严重拥堵即速度为 12km·h⁻¹ 时，交通量为 1892pcu·(h·ln)。见附件 A 表 A-1 所示，为连续时间段采集的实测数据和实时路况数据并用定量化值表达的对照表。

## 2.2 养护作业区延误计算以及开始时间优化

### 2.2.1 数据采集

在这个案例中，以上海市内环高架某路段处的养护作业开始时间优化进行演示。

由于没有现成的实时路况与速度定量化关系可供参考，本次研究对内环高架进行了标定。选择曹杨路段作为标定位置，该路段会出现常发性拥堵。同 2.1.1，在路段现场以及上游设置两个摄像机录制交通流视频，并用实时路况采集系统监测路段的实时路况变化，值得注意的是，这次数据采集用的实时路况地图是百度地图。

此外，养护作业开始时间优化需要养护路段的日常路况数据，本研究采用 1.1 中所介绍的系统，添加事件对要进行养护的位置进行实时路况监测，持续采集了一周该位置的实时路况数据，期间没有任何施工情况以及特殊节假日，最后取各天的平均路况作为该路段的日常交通情况，如图 8 所示。

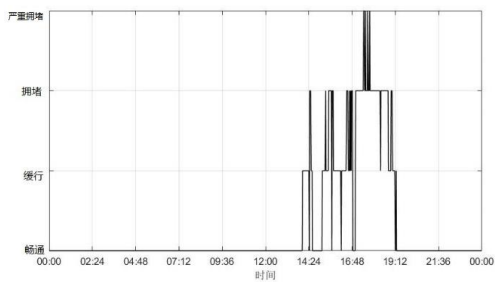

**图 8 实例养护作业位置处日常交通变化折线图**

Fig 8. Daily traffic status of the case study

### 2.2.2 实时路况转交通量

（1）实时路况转速度，方法同 2.1.2，对曹杨路段的交通流录像进行分析，得到车辆的速度，并结合实时路况数据，对不同路况的车辆速度进行分析，拟合成正态分布，取各状态曲线的交点作为各状态的分界点，并取各拟合正态曲线的均值作为该状态的定量指标，如下表 7 所示。

**表 7 实时路况状态定量表**

Table 7 Quantification values of the real-time traffic status of the case study

| 路况 | 畅通 | 缓行 | 拥堵 | 严重拥堵 |
|---|---|---|---|---|
| 速度范围(km·h⁻¹) | 52-80 | 30-52 | 11-30 | 0-11 |
| 定量化值(km·h⁻¹) | 62 | 44 | 18 | 7 |

由于养护路段的道路条件与曹杨路段一致，因此可以将曹杨路段的标定结果运用到养护路段[24]。有了上述实时路况与定量化速度的转换关系，就可以把养护作业位置处的路况数据转换为速度，得到如图 9 所示。

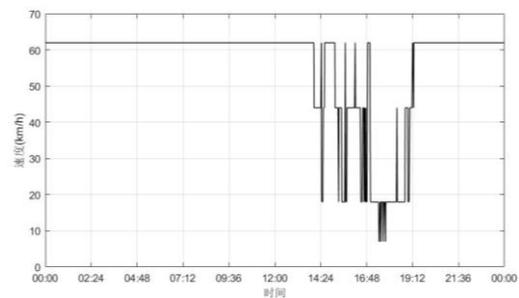

**图 9 实例养护作业位置处日常交通定量化后的折线图**

Fig 9. Daily traffic of the case study after quantification

（2）速度转交通量，首先要先得到养护作业路段的车速-流量实用关系模型。该路段单方向有 3 个车道，车道宽为 3.5m，左右路缘带都是 0.25m。基准自由流速度为 80km·h⁻¹，参考通行能力手册，对自由流修正后为 65 km·h⁻¹，确定不设养护作业区时单车道通行能力为 1500pcu·h⁻¹。

根据 1.3.1 所述的模型，对于本路段，取 $\alpha_1$=1.05，$\alpha_2$=1.88，$\alpha_3$=4.90，$U_s$=80。可以得到，当路况为顺畅即速度取 62km·h⁻¹ 时，交通量为 1170pcu·(h·ln)⁻¹；当路况为缓行即速度为 44 km·h⁻¹ 时，交通量为 1515pcu·(h·ln)⁻¹；当路况为拥堵即速度为 18km·h⁻¹ 时，交通量为 1695pcu·(h·ln)⁻¹；当路况为严重拥堵即速度为 7km·h⁻¹ 时，交通量为 1860 pcu·(h·ln)⁻¹。

将养护作业位置处的速度转换成交通量，得到如下图 10 所示的流量变化曲线。

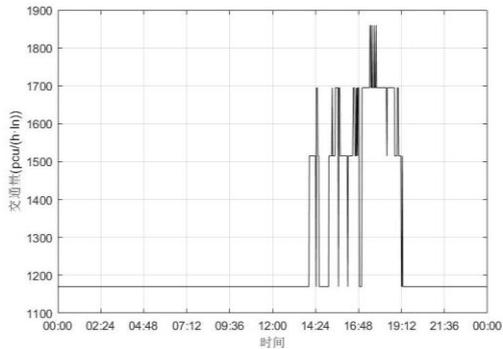

**图 10 养护作业位置实时交通转换成流量后的折线图**

Fig 10. Daily traffic volume of the case study

### 2.2.3 养护作业区通行能力估计

本次施工拟采用如图 11 所示的作业区布局。原来单向有三车道，每个车道 3.5m，现封闭一个车道施工；左侧向净空为 0.25，右侧侧向净空为 0；100%都是小车；施工区限速为 40km·h⁻¹；夜间施工。

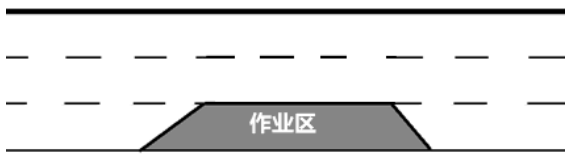

**图 11 实例拟采用的作业区布局**

Fig 11. Layout of the work zone

由 1.4 可得，该养护作业区基准通行能力为 1800 pcu·(h·ln)⁻¹。根据实际通行能力计算公式(5)，可以得到

$$C_{rs} = C_{bs} \times f_n \times f_{lw} \times f_{lc} \times f_{HV} \times f_{sc} \times f_{wi} \times f_{ls}$$
$$= 1800 \times 1 \times 0.97 \times (1 \times 0.96) \times 1 \times 0.8 \times 1 \times 0.96$$
$$= 1287 pcu \cdot (h \cdot ln)^{-1}.$$

### 2.2.4 养护作业区延误计算

现假定养护作业从晚上 8 点开始，即晚上 8 点养护作业区出现，养护作业持续时间为 8 小时，即凌晨 4 点时养护作业区撤去。根据 1.5 节介绍的方法计算得到总延误为 66885 小时。

### 2.2.5 养护作业开始时间优化

改变养护作业开始时间，比如将其调整到早上 6:00，以对比不同开始作业时间对总延误的影响，在条件允许时可以选择使总延误最小的开始时间开始养护施工。不同养护作业开始时间下的总延误如下图 12 所示。可以看到，在不同的时间开始施工对总延误有很大的影响。为了尽可能降低延误，最优的养护作业开始时间是晚上 9 点-11 点之间的任意时刻。

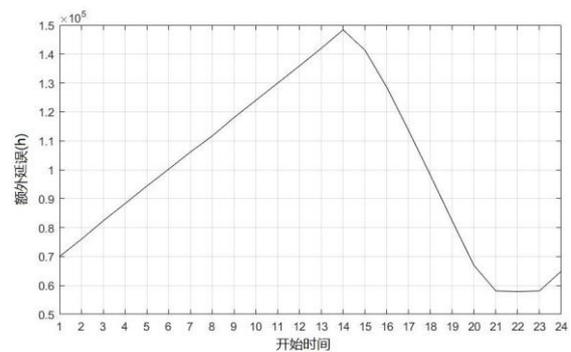

**图 12 以不同的时间开始施工的总延误变化图**

Fig 12. Total delay changing with different operation start time

## 3 讨论

前述章节对本研究的方法进行了介绍，并用实例进行了有效性验证。以下几点可做进一步讨

论：

（1）为了提高实时路况定量化的精度，有条件时应该对每条道路都进行一次标定。同一条道路，当道路条件（车道数、车道宽度和限速等）不发生改变时，定量标准可以认为是相同的；条件限制时，在保证道路条件相同的前提下，可以采用其他道路的标定结果。此外，由第 2 节的两个实例可以看到，选择不同的实时路况地图，路况与速度的转换关系也存在一定的不同，因此，即使是同一条道路，当路况地图更改时，也需要重新标定。

（2）本文分别在 1.3.2 和 1.4 两个小节对正常道路和养护作业区的通行能力估计进行了介绍，在有条件的情况下，建议通过实际数据检测得到真实的通行能力；如果条件不允许，再采用本文提到的通行能力确定方法。

（3）本方法在实时路况定量化过程中，顺畅状态有较大的误差：顺畅时，可能交通量很小，但案例中统一将交通量定为 1170pcu·(h·ln)$^{-1}$。尽管如此，这对于养护时间决策影响却很小，因为不管顺畅状态下实际的交通量为多少，相比于其他状态，该状态下的延误都是最小的。本文的处理方法效果上是对交通量进行保守估计，因此在应对实际交通流时还有一定的缓冲余地，对实际养护和道路出行是有利的。

## 4 结论与展望

本研究提出了一种基于实时路况地图数据的养护作业开始时间优化方法。首先提出了基于实时路况地图的实时路况数据采集方法，然后提出了将实时路况数据转换成实时交通量的方法，最后基于经典的排队论延误计算参数方法，实现对养护作业开始时间的优化。通过实际案例验证了实时路况定量化过程以及速度-流量实用关系模型的可靠性，验证了本文提出的将实时路况地图数据转换成实时交通量数据的可行性，也验证了基于实时路况地图实现对养护作业开始时间优化的有效性。本研究的一个重要贡献是，可以获取到有实时路况地图数据的道路网任意位置的实时交通量并进行积累，解决了广泛采用的基于排队论的延误计算参数法在实时路况数据采集上的困难，使得该方法可以更好的为养护决策和规划提供支撑，特别适用于养护作业历史数据不足或者不全的场合。

实时路况地图现在已经是电子地图非常常见的功能，其具有开放、更新及时、覆盖范围广等特点，可以为养护作业决策提供简单高效而成本低的数据来源，期待养护部门充分利用这一数据源，搭建相关的数据采集系统以及决策支持模块，优化养护作业的实际工作。

**参考文献：**